\documentclass[twocolumn,showpacs,amsmath,amssymb]{revtex4}

\usepackage[dvips]{graphics}
\usepackage{dcolumn}
\usepackage{bm}

\begin{document}

%\preprint{APS/123-QED}

\title{Kelvin modes of a fast rotating Bose-Einstein Condensate}

\author{F. Chevy$^{*}$}
\affiliation{Laboratoire Kastler Brossel$^{**}$,
 24 rue Lhomond 75005 Paris, France.
}

\author{S. Stringari}
 \affiliation{
Dipartimento di Fisica, Universit\`a di Trento and BEC-INFM,
I-38050 Povo, Italy.
}%

\date{\today}

\begin{abstract}

Using the concept of diffused vorticity and the formalism of rotational
hydrodynamics
we  calculate the eigenmodes of a harmonically trapped
Bose-Einstein condensate  containing an array of quantized vortices.  We predict the occurrence of a new branch of anomalous excitations, analogous to the
Kelvin modes  of the single vortex dynamics. Special attention is devoted to the excitation of the
anomalous scissors mode.

\end{abstract}


\maketitle

\subsection{Introduction}

The existence of the macroscopic wave function describing a
quantum fluid imposes a  velocity flow curl free everywhere except
on singularity lines known as vortices. Around these lines, the
circulation of the velocity is non zero and is quantized in unit
of $h/m$ where $m$ is the mass of the particles of the fluid. Recent
experiments have demonstrated the nucleation of such quantized
vortices in stirred gaseous Bose-Einstein condensates (BEC)
\cite{Matthews00,Madison00,AboShaer01,Hodby01}.

Multiple quantized vortices are energetically unstable  in
harmonic traps so that for large rotation frequencies,
Bose-Einstein condensates nucleate several singly quantized
vortices  that were observed to form regular triangular lattices
known as Abrikosov lattices \cite{AboShaer01,Madison00b,Haljan01}.
In this configuration, the circulation of the velocity field over
a circle orthogonal to the rotation axis and of radius $R$ much
larger than the vortex interspacing, is simply $\pi R^2n_vh/m$,
where $n_v$ is the density of the vortex lines. This is the same
formula as that of the velocity field of a rigid body rotating at
the  angular velocity

\begin{equation}
\Omega_0=n_v\frac{h}{2m}. \label{Eqn1}
\end{equation}

The dynamical properties of a single vortex line were first
studied by Lord Kelvin \cite{Thomson80} and  his results were
transposed to quantum fluids
\cite{Pitaevskii61,Isoshima99,Svidzinsky00}. For an excitation of
wave vector $k$ propagating along the rotation axis, the
dispersion relation  $\omega_{\rm K}$ of
these modes (Kelvin modes or kelvons) is:

\begin{equation}
\omega_{\rm K}\sim\frac{\hbar k^2}{2m}\ln \left(1/k\xi\right),
\label{Eqn1b}
\end{equation}

\noindent where $\xi=(8\pi\rho a)^{-1/2}$ is the healing length
giving the vortex core diameter, $a$ the scattering length
characterizing atom binary interactions and $\rho$ is the density
of the gas. These modes present the very peculiar feature that
they only exist with a single helicity as recently  proved in the
experiments of \cite{Bretin02}. Indeed, the Kelvin-Helmholtz
theorem that constrains the vorticity to move along with the fluid
imposes an angular momentum equal to $-\hbar$ to the Kelvin modes.

Although the problem of studying the dynamics of a vortex array
seems more involved at first sight, it is considerably
simplified for long wavelength perturbations. Indeed, in this case
a coarse grain averaging method permits to smooth the discrete
nature of vortices. In the case of a homogeneous condensate, it
was shown that excitations of wavevector  $ k$
propagating
transversally to the rotation axis satisfy the Tkatchenko
dispersion relation \cite{Tkachenko69,Baym83}:

$$\omega^2_{\rm T}=\frac{\hbar\Omega_0}{4m}k^2,$$

\noindent where $\Omega_0$ is the effective  angular
frequency of the condensate defined by (\ref{Eqn1}). The Tkatchenko  modes are  elastic excitations of the lattice
 and
have been recently investigated theoretically also in the presence
of harmonic traps \cite{Anglin02}. First experimental evidences
for such modes has been reported in \cite{jila03}.

In this paper we use the coarse grain method to  study
the eigenmodes of a  rotating Bose-Einstein
condensate confined by a harmonic trap using a fully hydrodynamic approach including vorticity
\cite{Sedrakian01}. In addition to the usual collective modes exhibited by the condensate
in the absence of rotation, we identify an additional branch, analogous to the Kelvin modes exhibited by a single vortex line.  With respect to the Tkatchenko modes, whose frequencies
 vanish in the Thomas Fermi limit \cite{Anglin02}, the
Kelvin  excitations  emerging
from our hydrodynamic picture approach a finite value in the Thomas-Fermi limit. In the case of elongated
traps their frequencies actually  scale like $\omega^2_z/\omega_{\perp}$, for a fixed value of $\Omega_0/\omega_{\perp}$.

The paper is organized as follows: in Sect. B we develop the
formalism of rotational hydrodynamics in the presence of harmonic
trapping and derive the general equations (\ref{dispersion}) for
the dispersion law of the linearized excitations. In Sect. C we
briefely summarize the results for the surface excitations which
represent a natural generalization of the modes exhibited by non
rotating Bose-Enstein condensates.  Section D is devoted to the
scissors mode where an anomalous mode of Kelvin nature is
predicted. The properties of the scissors oscillations for a
rotating condensate are discussed in  detail using the formalism
of linear response theory. Finally in Sect. E we obtain a general
dispersion relation for the Kelvin modes in  elongated traps.

\subsection{Rotational hydrodynamics and elementary excitations}

It is well known that in the
so-called Thomas-Fermi regime, where the  mean field interaction
dominates over the quantum pressure, the dynamics of a non rotating condensate
can be described by the  classical equations of hydrodynamics:

\begin{eqnarray}
\partial_t\rho&=&-\mathbf{\nabla}\left(\rho \mathbf{v}\right)\label{hydroa}\\
m\partial_t\mathbf{v}&=&-\mathbf{\nabla}\left(U_t+g\rho+mv^2/2\right),
\label{hydrob}
\end{eqnarray}

\noindent where $\rho$ is the local particle density,
$g=4\pi\hbar^2a/m$ is the coupling constant characterizing the interatomic force
and $\mathbf{v}$ the velocity field satisfying the irrotational condition
$\mathbf{\nabla}\times\mathbf{v}=0$.
In the case of  cylindrical
  harmonic trapping,

$$U_t=\frac{m}{2}\left(\omega_\perp^2 (x^2+y^2)+\omega_z^2
z^2\right) \; ,$$
 the hydrodynamic equations (\ref{hydroa}) and (\ref{hydrob})  admit a class of  analytic solutions
 \cite{sandro96} whose  frequencies have been confirmed experimentally with
high accuracy.

 In the presence of
vortex lines the hydrodynamical formalism must be modified. Here
we will employ a simplified procedure, by assuming that the
characteristic wavelength of excitations is large enough so that
we can average all the physical quantities over domains containing
several vortices. In this case the average lab frame velocity
field $\mathbf{\bar v}$ is no longer curl-free. Since each vortex
carries a flux $h/m$, the average vorticity of the flow is:

\begin{equation}
\mathbf{\bar\Omega}=\frac{\mathbf{\nabla\times\bar
v}}{2}=n_v\frac{h}{2m}\mathbf{u}, \label{Eqn2}
\end{equation}

\noindent where $n_v$ is the vortex surface density and the unit vector
$\mathbf{u}$ is the local direction of the vortex lines. According
to equation (\ref{Eqn2}), the average vorticity
$\mathbf{\bar\Omega}$ characterizes the local vortex distribution:
its direction indicates the orientation of  the vortex lines while its
modulus is proportional to the vortex density.

Like in \cite{Sedrakian01,Cozzini02}, we shall  simply
assume that the average velocity field and density satisfy
classical hydrodynamical equations, including rotational terms,
namely:

\begin{equation}
\left\{
\begin{array}{rcl}
\partial_t\bar\rho&=&-\mathbf{\nabla}\left(\bar\rho \mathbf{\bar v}\right)\\
m\partial_t\mathbf{\bar
v}&=&-\mathbf{\nabla}\left(U_t+g\bar\rho+m\frac{\bar
v^2}{2}\right)-m\left(\mathbf{\nabla\times\bar
v}\right)\times\mathbf{\bar v},
\end{array}
\right. \label{Eqn3}
\end{equation}

A stationary solutions of equations (\ref{Eqn3}) is given by:

\begin{equation}
\left\{
\begin{array}{rcl}
\mathbf{\bar v_0}&=&\mathbf{\Omega_0\times r}\\
g\bar\rho_0/m&=&\mu-\left(\omega_\perp^2-\Omega_0^2\right)(x^2+y^2)/2-\omega_z^2z^2/2,
\end{array}
\right. \label{Eqn4}
\end{equation}

\noindent where $\mu$ is the chemical potential and
$\mathbf{\Omega_0}=\Omega_0{\bf u}_z$ characterizes the vortex
density and direction according to equation (\ref{Eqn2}). As
expected, the stationary velocity field is equivalent to that of a
rigid body rotating at  angular velocity $\mathbf{\Omega}_0$,
while the density is described by the usual Thomas-Fermi profile
with the trapping potential $U_t$  corrected by the centrifugal
potential $U_c=-m\Omega_0^2 (x^2+y^2)/2$.
The static behavior described above has been
verified experimentally. In particular, the modification of the
transverse trapping is used to measure experimentally the
effective angular velocity  $\mathbf{\Omega_0}$ \cite{AboShaer01,Haljan01}. More
interesting effects concern the study of the dynamics of the condensate. In \cite{Cozzini02} the hydrodynamic equations have been used to study the time evolution of a condensate containing a vortex array, following the sudden switch on
of a static deformation in the plane of rotation. This produces peculiar non linear effects that have been experimentally observed in \cite{Haljan01}.  In the following we will focus on the behaviour of the linearized solutions
which can be derived by looking for small  perturbations of the
the density and velocity field:

$$\left\{
\begin{array}{rcl}
\bar\rho&=&\bar\rho_0+\delta\rho\\
\mathbf{\bar v}&=&\mathbf{\Omega_0\times r}+\delta\mathbf{v},
\end{array}
\right.
$$

\noindent with respect to the equilibrium values (in this case,
$\delta {\bf v}$ can be interpreted as the velocity in the
rotating frame). If the characteristic wavelength of the
perturbation is much larger than the vortex spacing, the
``averaged'' hydrodynamical equations (\ref{Eqn3}) can still be
applied and, in the {\em frame rotating at the angular velocity}
$\mathbf{\Omega_0}$, they read, in linear approximation,

\begin{eqnarray}
\partial_t\delta\rho&=&\mathbf{\nabla'}\left(\bar\rho_0\delta\mathbf{v}\right)
\label{Eqn5}\\
\partial_t\delta\mathbf{
v}&=&-\mathbf{\nabla'}\left(\frac{g\delta\rho}{m}\right)-2\mathbf{\Omega}_0\times\delta\mathbf{v},
\label{Eqn6}
\end{eqnarray}

\noindent where $\mathbf{\nabla}'$ denotes the derivation with
respect to the coordinates in the rotating frame. The system of
equations (\ref{Eqn5},\ref{Eqn6}) constitutes the starting point
of our analysis. We will study its eigenmodes and show that the
spectrum, in addition to the usual ``phonon" modes displays a new
class of solutions carrying negative angular momentum and that can
be physically regarded as
 Kelvin excitations.

Let us  set $\partial_t=-i\omega'$ in equations
(\ref{Eqn5},\ref{Eqn6}). Using equation (\ref{Eqn6}), we can then
express $\delta\mathbf{\bar v}$ as a function of $\delta\rho$ as

$$\delta\mathbf{v}=\frac{1}{\omega'^2-4\Omega_0^2}\left[i\omega'\left({\bf f}\right)_\perp+2\mathbf{\Omega}_0\times{\bf
f}\right]-\frac{1}{i\omega'}\left({\bf f}\right)_\|,$$

\noindent where ${\bf
f}=-\mathbf{\nabla}'\left(g\delta\rho\right)/m$ and $\left({\bf f}\right)_\|$ and $\left({\bf
f}\right)_\perp$ are, respectively, the projections  of ${\bf f}$ on the
 directions parallel and orthogonal to $z$.

Inserting this expression for the velocity field in the linearized
mass conservation equation (\ref{Eqn5}) , we get a closed equation
for the density fluctuations that can be written as

\begin{equation}
i\omega'\left(\omega'\null^2-4\Omega_0^2\right)\delta\rho={\cal
A}\cdot\delta \rho,
\label{dispersion}
\end{equation}

\noindent where ${\cal A}$
  denotes the linear
operator

\begin{eqnarray*}{\cal
A}\cdot\delta\rho&=&-\nabla'_\perp\left[\frac{g\rho_0}{m}\left(i\omega'\nabla'_\perp\delta\rho+2\mathbf{\Omega_0}\times\nabla'_\perp\delta\rho\right)\right]\\
&&+\nabla'_\|\left[\frac{g\rho_0}{m}\left(\frac{\omega'\null^2-4\Omega_0^2}{i\omega'}\nabla'_\|\delta\rho\right)\right],
\label{A}
\end{eqnarray*}
and  $\rho_0$ is the equilibrium density profile  (\ref{Eqn4}).
Equations (\ref{dispersion}) and (\ref{A}) are the main result of
this paper. Simple solutions  can be found in the form of
polynomials.

  Let us  stress again
that the calculation of $\omega'$ is performed in the {\em
rotating frame}. The corresponding frequency $\omega$ in the lab
frame is obtained  through the simple relation
$\omega'=\omega-m_z\Omega_0$, where $\hbar m_z$ is the
angular momentum of the excitation along the rotation axis. Since
both $\delta\rho$ and its complex conjugate are solutions
of (\ref{A})  we find that for any solution with
angular momentum $m_z$ and eigenfrequency $\omega'$, there should
be an other solution with angular momentum $-m_z$ and frequency
$-\omega'$. Since the unperturbed state (condensate rotating
at angular velocity  $\mathbf{\Omega_0}$) is the ground state of the system in the
rotating frame, the physical solutions  carrying energy $\hbar \omega'$ should correspond
to positive frequencies $\omega'$.

Note also that the equation (\ref{dispersion}) is valid for
$\omega'\not= 0$ and $\omega'^2\not = 4\Omega_0^2$ and that these
two special cases must be treated separately. The case
$\omega'^2-4\Omega_0^2=0$ does not lead to any new mode. The case
$\omega'=0$ is instead more interesting since it is related to the
Tkatchenko's modes, as  stressed in the introduction. Starting the
analysis back from equations (\ref{Eqn5}) and (\ref{Eqn6}), we can
show that the density perturbation $\delta\rho$ of a zero energy
modes
 depends only on the radial coordinate $r_\perp=\sqrt{x^2+y^2}$ and that the velocity field is given by:

$$\left\{
\begin{array}{rcl}
v_z&=&0\\
{\bf
v}_\perp&=&\displaystyle{\frac{g}{2m\Omega_0^2}}\,\mathbf{\Omega}_0\times\mathbf{\nabla}\delta\rho.
\end{array} \right.
$$

In the case where $\delta\rho$ is parabolic, this zero energy mode
can be simply interpreted as a modification of $\Omega_0$, {\em
i.e.} a change of the vortex density.

The rest of the paper is devoted to the solution of equation
(\ref{A}) for special cases of physical interest. We will first
discuss  the  surface oscillations then the scissors modes, that
will be shown to exhibit a kelvon-like branch. Finally we will
derive the general dispersion law for the Kelvon modes working in
the geometry of elongated traps.

\subsection{Surface modes}

Surface modes are characterized by the form $\delta\rho=(x\pm
iy)^l$ and carry angular momentum $m_z=\pm l$. Inserting this
Ansatz in the differential equation (\ref{dispersion})  yields the
following equation for the eigenfrequencies $\omega'_{\pm l}$ in
the rotating frame:

$$\omega'_{\pm l}\null^2\pm 2\Omega_0\omega'_{\pm l}-l
\left(\omega_\perp^2-\Omega_0^2\right)=0.$$

The positive solution of this equation  reads:

$$\omega'_{\pm
l}=\sqrt{l\omega_\perp^2 -(l-1)\Omega_0^2} \mp\Omega_0 ,$$

\noindent or, in the laboratory frame,

$$\omega_{\pm l}=\sqrt{l\omega_\perp^2 -(l-1)\Omega_0^2} \pm(l-1)\Omega_0 .
$$

Contrarily to what happens in the case of a non-rotating
condensate, the two modes are no longer degenerate, and the
degeneracy lift amounts to:

\begin{equation}
\Delta\omega_l=\omega_{+l}-\omega_{-l}=2
(l-1)\Omega_0. \label{Eqn8}
\end{equation}

This result is valid for any $l$ \cite{LargeL} but can be checked
for some special cases previously reported in the literature.

In the  $l=1$ case (dipole motion) one finds $\Delta \omega=0$. In this case
the eigenfrequencies are actually unaffected by the rotation
since $\omega_{\pm l}=\omega_\perp$ irrespectively of the value of
$\Omega_0$. This is not surprising since we know that the
generalized Kohn theorem \cite{Dobson94} implies that the center
of mass motion is determined only by the frequency of the harmonic trap.

The case $l=2$ is also well documented, both theoretically and
experimentally \cite{Zambelli01,Chevy00}. Using sum-rule approach,
it can be shown in particular that the degeneracy lift amounts to:

\begin{equation}
\Delta\omega=2\frac{\langle \ell_z\rangle}{m\langle
r_\perp^2\rangle}, \label{Eqn9}
\end{equation}

\noindent $\ell_z$ is the  angular momentum per particle along $z$
of the unperturbed condensate. Since the unperturbed velocity
field is rigid-like, the angular momentum is given by the
classical formula $\langle \ell_z\rangle=m\langle r_\perp^2\rangle
\Omega_0$. Inserting this relation in (\ref{Eqn9}) yields result
(\ref{Eqn8}) for $l=2$.

\begin{figure}[t]
\centerline{\scalebox{.95}{\includegraphics{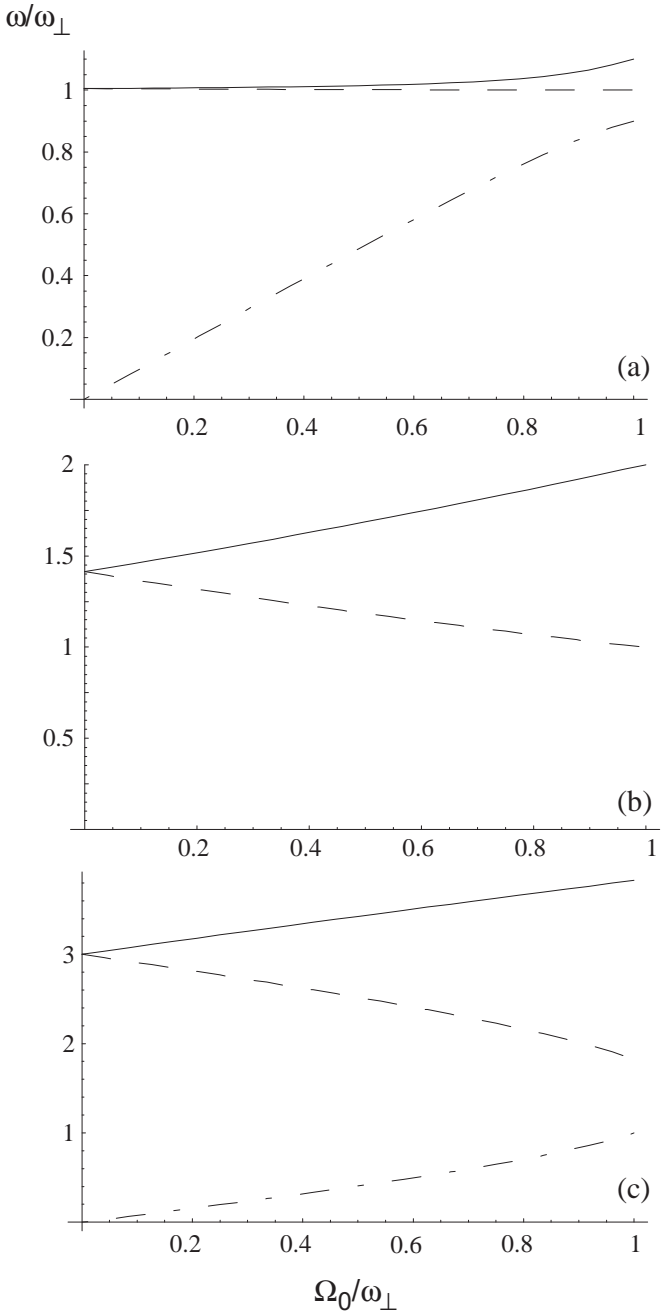}}}
 \caption{Absolute values
of the lab frame frequencies  of the  $m_z=+1$ (full line),
$m_z=-1$ (dashed line) and anomalous (dash-dotted line) scissors
modes for three trap geometries. (a): cigar shaped, with
$\omega_\perp=10\omega_z$. (b): isotropic trap with
$\omega_z=\omega_\perp$. In this case the frequency  of the
anomalous mode is zero. (c): pancake trap with
$\omega_z=\sqrt{8}\omega_\perp$.} \label{Fig1}
\end{figure}

\subsection{ Scissors modes}

 The scissors  modes are associated with a
density fluctuation $\delta\rho=(x\pm iy)z$ carrying angular
momentum $m_z=\pm 1$ along the $z$ axis. Using equation (\ref{A}),
one finds that  the eigenfrequencies  $\omega'$ must satisfy the
non trivial cubic equation:

\begin{equation}
\omega'\null^3\pm
2\Omega_0\omega'^2-\omega'\left(\omega_z^2+\omega^2_\perp-\Omega_0^2\right)\mp
2\Omega_0\omega_z^2=0. \label{Eqn10}
\end{equation}

This equation is analogous to that found by the authors of
\cite{Sedrakian01} using the tensor virial method. Two of the
resulting modes can be identified with the usual $m_z=\pm 1$
scissors previously studied in the case of non-rotating
Bose-Einstein condensates \cite{dgo}. The splitting between these
two modes is easily calculated when $\Omega_0 \to 0$. In the
laboratory frame one finds $\Delta \omega = 2\Omega_0
\omega_z^2/(\omega_z^2+\omega_{\perp}^2)$  in agreement with the
sum rule result of \cite{Zambelli01} that was confirmed
experimentally in \cite{Hodby02}. This splitting is at the origin
of the gyroscopic effect investigated in \cite{Stringari01} in the
presence of a single vortex line.
 The  third mode predicted by (\ref{Eqn10})  has instead  no analog  for $\Omega_0=0$. A sign
analysis shows that the frequency $\omega'_a$ of this {\em
anomalous} mode is positive for $m_z=-1$. Just like the kelvons,
this new scissor mode can only exist with negative helicity. The
analog of this mode in the case of a single vortex line was
investigated by \cite{Svidzinsky00}.  The solutions of equation
(\ref{Eqn10}) as a function of $\Omega_0$ are shown in Fig.1 for
different trapping geometries. The interpretation of this mode is
straightforward for an isotropic trap: in this case, the anomalous
mode is associated with an overall rotation of the condensate and
of its lattice. In an isotropic trap this rotation indeed costs no
energy and the frequency in the lab frame is exactly  zero.

For very deformed traps, {\em i.e.} for
$|\omega_\perp^2-\omega_z^2|\gg\Omega_0^2$, the anomalous mode is
associated to a scissors mode of the vortex lattice while the
density profile remains almost at rest.

In the second part of this section we discuss how the scissors
modes, and in particular the anomalous mode, can be excited and
observed  by suddenly tilting the trapping potential in the
($x,z$) plane. To this purpose we make use of the formalism of
linear response theory.  Let us  apply a perturbing potential of
the form

$$\delta U=\epsilon z'(x'+i y') e^{-i\omega' t}+\epsilon^* z' (x'-i y')
e^{i\omega' t}$$
in the rotating frame.
Using the hydrodynamical equations, we find that
the density and the vorticity will be perturbed as

\begin{eqnarray*}\delta\rho&=&\epsilon\lambda_{\omega'} z' (x'+iy') e^{-i\omega'
t}+\epsilon^*\lambda_{\omega'}^* z' (x'-i y') e^{i\omega' t},\\
\delta\mathbf{\Omega}&=&\epsilon\mu_{\omega'} e^{-i\omega' t}{\bf
u}_+ +\epsilon^*\mu_{\omega'}^* e^{i\omega' t}{\bf u}_-,
\end{eqnarray*}

\noindent where ${\bf u}_{\pm}={\bf u}_x\pm i{\bf u}_y$. The quantities $\lambda$
and $\mu$ can be calculated using the hydrodynamical equations and one finds the result

\begin{eqnarray}
\lambda_{\omega'}&=&\frac{2\Omega_0\omega_z^2+\omega'
(\omega_\perp^2+\omega_z^2-\Omega_0^2)}{g P(\omega')}\label{DensityResponse}\\
\mu_{\omega'}&=&-\frac{\omega'\Omega_0}{m P(\omega')},
\label{LatticeResponse}
\end{eqnarray}

\noindent where
$P(\omega')={\omega'}^3+2\Omega_0{\omega'}^2-\omega'
(\omega_z^2+\omega_\perp^2-\Omega_0^2)-2\Omega_0\omega_z^2$ is a
polynomial whose  roots fix the eigenfrequencies of the system
(see Eq.(\ref{Eqn10})). Notice that the change  $\delta {\bf
\Omega}_0$ in the vorticity, lying in the ($x,y$) plane, actually
corresponds to a rotation of the vector lattice. This rotation
could be imaged experimentally, allowing, together with the
changes in the shape of the atomic cloud, for a direct
identification of the various scissors modes.

On the one hand, the expectation value $\langle z'(x'-iy')\rangle
\equiv \int {\rm d}^3 r \delta\rho z'(x'-iy')$ can be written as

$$\langle z'(x'-iy')\rangle =
\epsilon\chi_\omega^{\rm HD}e^{-i\omega'
t}+\epsilon^*\left(\chi_\omega^{\rm HD}\right)^*e^{-i\omega't},$$
and one obtains the result

\begin{equation}
\chi_{\omega'}^{\rm HD}=\lambda_{\omega'} \frac{\pi R_z^3 R_\perp^4}{6}.
\end{equation}
for the hydrodynamic response function
$\chi_\omega^{\rm HD}$.

 On the other
hand, this expectation value can be calculated using the formalism of  quantum mechanics. Using first order
perturbation theory
the linear response function relative to the operator
$F=\sum_{k=1}^{N} z'_k (x'_k-iy'_k)$
can be in fact written as

\begin{equation}
\label{LinearQ} \chi_{\omega'}^{\rm QM}=\sum_n \frac{|\langle
n|F^\dagger|0\rangle|^2}{\hbar\omega'-E'_n}-\frac{|\langle
n|F|0\rangle|^2}{\hbar\omega'+E'_n},
\end{equation}

\noindent where $|0\rangle$ and $|n\rangle$ are respectively the
ground state and the excited states of  the many body system and
$E'_n$ are the associated  eigenenergies in the rotating frame.
From (\ref{LinearQ}), we see that the poles of $\chi_\omega$ are
the eigenfrequencies of the system. Moreover, the sign of the
residue at  the pole gives its helicity: positive residues are
associated with positive helicity ({\em i.e.} with modes excited
by $z (x+iy)$ ), and negative residues are instead associated
with negative helicity.

\begin{figure}[t]
{\includegraphics{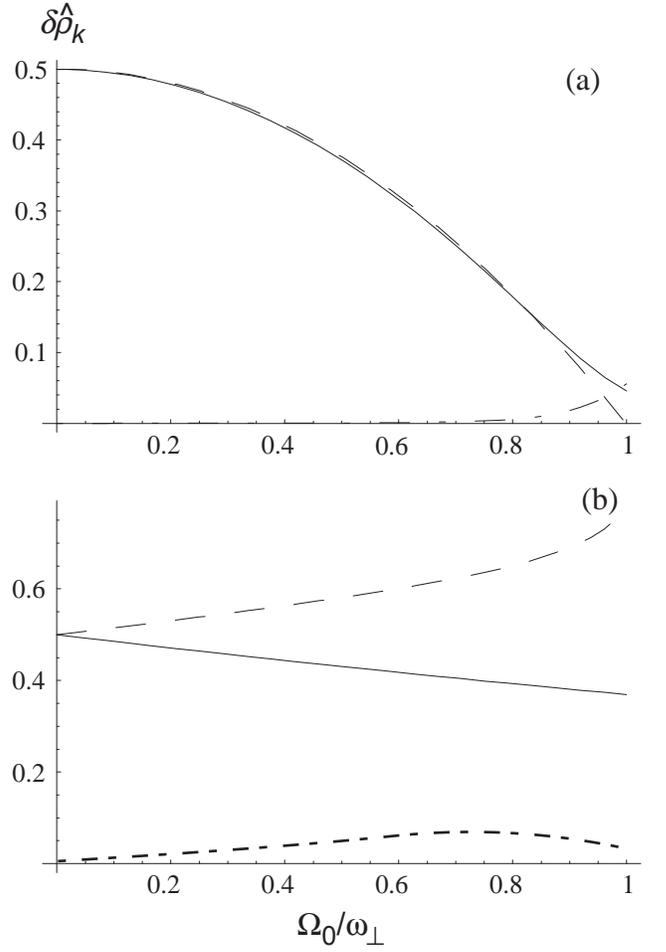}} \caption{Density coupling of the
scissor $m_z=+1$ (full line), $m_z=-1$ (dashed line) and anomalous
(dash dotted line) transitions for different trapping geometries.
(a) Elongated trap  ($\omega_\perp=10\omega_z$) and  (b) pancake
trap ($\omega_z=\sqrt{8}\omega_\perp$).
 )}
 \label{Fig2}
\end{figure}

By identifying $\chi^{\rm QM}$ and $\chi^{\rm HD}$ and by close
examination of eqs. (\ref{DensityResponse}),  we can conclude
that~:

\begin{enumerate}

\item The anomalous mode has positive energy  in the rotating
frame and its  angular momentum is  $m=-1$.

\item For $\omega_z<\omega_\perp$, we have $\omega_{\rm a}<0$~. For cigar geometries the
anomalous mode is then associated to a thermodynamical instability
of the vortex lattice in the laboratory frame \cite{Reverse}.

\item On the contrary, for $\omega_z>\omega_\perp$, we have
$\omega_{\rm a}>0$. For pancake geometries, the vortex lattice is
stable versus scissor perturbations.

\end{enumerate}

\begin{figure}[t]
{\includegraphics{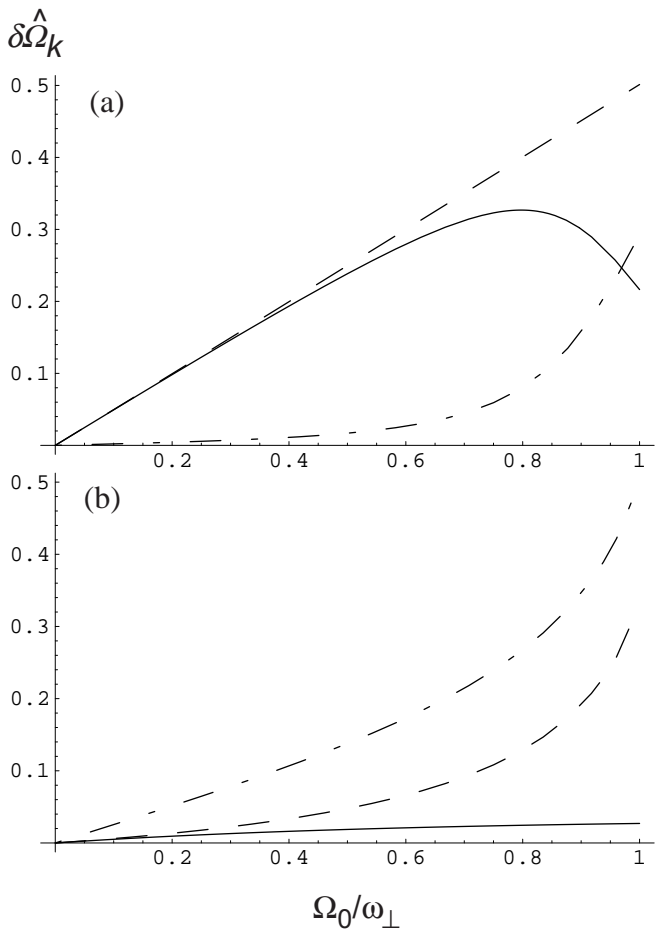}} \caption{Change of vorticity  of the
scissor $m_z=+1$ (full line), $m_z=-1$ (dashed line) and anomalous
(dash dotted line) transitions for different trapping geometries.
(a) Elongated trap  ($\omega_\perp=10\omega_z$) and (b) pancake
trap ($\omega_z=\sqrt{8}\omega_\perp$). For these modes the change
of vorticity is associated with a change of direction of the
vortical lattice.
 )}
 \label{Fig3}
\end{figure}

Let us now consider the special case of the sudden tilting at
$t=0$ of the longitudinal axis of the trap
 in the
lab frame. This excitation leads to the following perturbing
potential:

$$\delta U=\alpha xz Y(t)=\alpha z(x+iy)+\alpha z(x-iy) Y(t),$$

\noindent where $Y$ is the Heavyside step-function equal to $0$
for $t<0$ and 1 otherwise. Expressed in the rotating frame,
$\delta U$ reads:

$$\delta U=\left(\alpha z(x+iy)e^{i\Omega_0 t}+\alpha z(x-iy)e^{-i\Omega_0 t}\right)Y(t).$$

The Fourier transform of this perturbation then yields:

$$\epsilon_{\omega'}=\alpha(\pi\delta
(\omega'+\Omega_0)+i/(\omega'+\Omega_0)).$$

According to equation (\ref{DensityResponse}) and
(\ref{LatticeResponse}), the induced density and vorticity changes are given by:

\begin{eqnarray}
\delta\rho&=&\left(\int{\rm
d}\omega'\epsilon_{\omega'}\lambda_{\omega'}e^{-i\omega'
t}\right)z(x+iy)+{\rm c.c.}\label{Residue1}\\
\delta\mathbf{\Omega}&=&\left(\int{\rm
d}\omega'\epsilon_{\omega'}\mu_{\omega'}e^{-i\omega' t}\right){\bf
u}_++{\rm c.c.}.\label{Residue2}
\end{eqnarray}

These integrals can be calculated using  standard integration
techniques in the complex plane. In particular, $\delta\rho$ and
$\delta\mathbf{\Omega}$
 present terms oscillating at  the driving frequency
$\-\Omega_0$, as well as at the scissor frequencies $\omega_{\pm 1}$
and $\omega_{\rm a}$. Let us introduce the Fourier components of
these modes:

\begin{eqnarray*}
\delta\rho&=&\left(\delta\hat\rho_0 e^{i\Omega_0 t}+\sum_{k=\pm 1,a}\delta\hat\rho_k e^{-i\omega'_k t}\right)z(x+iy)+{\rm c.c.}\\
\delta\mathbf{\Omega}&=&\left(\delta\hat\Omega_0 e^{i\Omega_0
t}+\sum_{k=\pm 1,a}\delta\hat\Omega_k e^{-i\omega'_k t}\right){\bf
u}_++{\rm c.c.}.
\end{eqnarray*}

According to equations (\ref{Residue1}) and (\ref{Residue2}),
$\delta\hat\rho_k$ and $\delta\hat\Omega_k$ are given, respectively,
by the residues of $\lambda_{\omega'}/(\omega'+\Omega_0)$ and
$\mu_{\omega'}/(\omega'+\Omega_0)$ . We have plotted the density
response $\delta\hat\rho_k$ on Fig. (\ref{Fig2}). We see that the
anomalous mode weight is very weak, both in the pancake and cigar
traps (except for $\Omega_0\sim\omega_\perp$). The behavior of the
vorticity is dramatically different, as observed on Fig.
(\ref{Fig3}). Indeed, while the anomalous mode remains weak in the
case of an elongated trap (Fig. (\ref{Fig3}.a)), we see that it
dominates the dynamics in the pancake geometry. This effect should be detectable
experimentally by imaging the orientation of the vortex lattice.

\subsection{Kelvin spectrum}

 The study of the scissor
modes has revealed the existence of an anomalous kelvon-like mode with
angular momentum $m_z=-1$. For very elongated traps
($\omega_\perp\gg\omega_z$), the corresponding solution is given by:

$$\omega'\sim
2\Omega_0\frac{\omega_z^2}{\omega_\perp^2-\Omega_0^2}$$ showing
that the   frequency of the anomalous mode goes to zero in the
limit of small $\omega_z$. In what follows, we shall generalize
this result by looking for more general solutions of (\ref{Eqn8})
satisfying $\omega'\sim\omega_z^2$ when $\omega_z\rightarrow 0$.
In this approximation, we can restrict equation (\ref{A}) to non
vanishing terms in $\omega$ and $\omega_z$, which yields the
simplified equation

\begin{equation}
\nabla'_\perp\left[2\rho_0
\mathbf{\Omega_0}\times\nabla'_\perp\delta\rho\right]+\nabla'_\|\left[\rho_0
\left(\frac{4\Omega_0^2}{i\omega'}\nabla'_\|\delta\rho\right)\right]=0.
\label{Eqn11}
\end{equation}

Let us now  write the density perturbation as the most general
 polynomial of order $p+l$ associated to angular momentum
$m_z=-l$:

$$\delta\rho=r_\perp^{l}e^{-il\theta}\sum_{q=0}^{q_m}b_qz^{2q}r_\perp^{p-2q}+...,$$

\noindent where the expression is restricted to terms of leading
order in the  $(r_\perp,\theta,z)$ cylindrical coordinates and
where $q_m$ is the highest $q$ such that $p-2q\ge 0$. According to
equation (\ref{Eqn11}), the coefficients $b_{q>0}$ must satisfy
the
 recursive relation:

$$b_q
=-\frac{(p-2q+2)(p+1-2q)\Omega_0^2}{\left(-l\Omega_0\omega_\perp^2\omega'+(p-2q)(p-2q+1)\Omega_0^2\omega_z^2\right)}b_{q-1},$$

\noindent while, for $q=0$,

\begin{equation}\left(-l\Omega_0\omega_\perp^2+p(p+1)\frac{\Omega_0^2}{\omega'}\omega_z^2\right)b_0=0.
\label{Eqn12}
\end{equation}

In order to get non vanishing $b_q$, the coefficient of $b_0$ in
equation (\ref{Eqn12}) must cancel out. This ensures the
quantization of the eigenfrequencies, since we must have:

\begin{equation}
\omega'=p(p+1)\frac{\Omega_0\omega_z^2}{l\left(\omega_\perp^2-\Omega_0^2\right)}.
\label{Eqn13}
\end{equation}

This dispersion relation represents  the generalization of the
classical Kelvin law to the case of a rotating gas confined in a
harmonic trap. Like kelvons, these new modes have always a
negative angular momentum and possess a quadratic behavior for
large $p$. If $R_z$ and $R_\perp$ denote the condensate radii in
the longitudinal and transverse directions respectively,  these
two quantities are related by (see (\ref{Eqn4}))

$$\omega_z^2R_z^2=\left(\omega_\perp^2-\Omega^2\right)R_\perp^2.$$

Moreover,  $\delta\rho$ is a polynomial of order $p$ in $z$. The
quantity $k= p/R_z$ can then be interpreted as the longitudinal
wave vector of the excitation. Using the relation $\Omega_0=n_v
h/2m$ equation (\ref{Eqn13}) yields, for large quantum numbers
$p$,

$$\omega'=N_v\frac{\hbar}{lm}k^2,$$

\noindent where $N_v=n_v \pi R_\perp^2$ is the number of vortices
present in the condensate. This dispersion law is very similar to
equation (\ref{Eqn1b}), except for the logarithmic factor that
vanished through the averaging procedure.

Note also that  equation (\ref{Eqn13}) is singular for $l=0$. This
is due to the fact that in this case $\omega'\sim\omega_z$ so
terms neglected in (\ref{Eqn11}) must be taken into account. In
this case, there is no decoupling between the phonon and kelvon
branches, and no simple analytical expression can be extracted.

\begin{acknowledgments}
We wish to acknowledge M. Cozzini, L. Pitaevskii and G. Baym for
very helpful discussions.
\end{acknowledgments}


\begin{thebibliography}{30}
\bibitem{LargeL} $l$ must be compatible with the large wavelentgh
hypothesis, that is $2\pi R_\perp/l$ should be much larger than
the vortex interspacing.
\end{thebibliography}
\end{document}